
  \font\twelverm=cmr10 scaled 1200       \font\twelvei=cmmi10 scaled 1200
  \font\twelvesy=cmsy10 scaled 1200      \font\twelveex=cmex10 scaled 1200
  \font\twelvebf=cmbx10 scaled 1200      \font\twelvesl=cmsl10 scaled 1200
  \font\twelvett=cmtt10 scaled 1200      \font\twelveit=cmti10 scaled 1200

  \font\twelvemib=cmmib10 scaled 1200
  \font\elevenmib=cmmib10 scaled 1095
  \font\tenmib=cmmib10
  \font\eightmib=cmmib10 scaled 800


\font\elevenrm=cmr10 scaled 1095    \font\eleveni=cmmi10 scaled 1095
\font\elevensy=cmsy10 scaled 1095


\font\seventeeni=cmmi10 scaled \magstep3

\font\seventeensy=cmsy10 scaled \magstep3

\font\seventeenmib=cmmib10 scaled \magstep3

\newfam\cpfam%


 \font\eightbf=cmbx8


\skewchar\eleveni='177   \skewchar\elevensy='60
\skewchar\elevenmib='177  \skewchar\seventeensy='60
\skewchar\seventeenmib='177
\skewchar\seventeeni='177

\newfam\mibfam%


  \skewchar\twelvei='177   \skewchar\twelvesy='60
  \skewchar\twelvemib='177
%
%
\def\twelvepoint{\normalbaselineskip=12.4pt
  \abovedisplayskip 12.4pt plus 3pt minus 9pt
  \belowdisplayskip 12.4pt plus 3pt minus 9pt
  \abovedisplayshortskip 0pt plus 3pt
  \belowdisplayshortskip 7.2pt plus 3pt minus 4pt
  \smallskipamount=3.6pt plus 1.2pt minus 1.2pt
  \medskipamount=7.2pt plus 2.4pt minus 2.4pt
  \bigskipamount=14.4pt plus 4.8pt minus 4.8pt
  \def\rm{\fam0\twelverm}          \def\it{\fam\itfam\twelveit}%
  \def\sl{\fam\slfam\twelvesl}     \def\bf{\fam\bffam\twelvebf}%
  \def\mit{\fam 1}                 \def\cal{\fam 2}%
  \def\tt{\twelvett}%
  \def\mib{\fam\mibfam\twelvemib}%

  \textfont0=\twelverm   \scriptfont0=\tenrm     \scriptscriptfont0=\sevenrm
  \textfont1=\twelvei    \scriptfont1=\teni      \scriptscriptfont1=\seveni
  \textfont2=\twelvesy   \scriptfont2=\tensy     \scriptscriptfont2=\sevensy
  \textfont3=\twelveex   \scriptfont3=\twelveex  \scriptscriptfont3=\twelveex
  \textfont\itfam=\twelveit
  \textfont\slfam=\twelvesl
  \textfont\bffam=\twelvebf \scriptfont\bffam=\tenbf
                             \scriptscriptfont\bffam=\eightbf
  \textfont\mibfam=\twelvemib       \scriptfont\mibfam=\tenmib
                               	      \scriptscriptfont\mibfam=\eightmib

  \def\xrm{\textfont0=\twelverm\scriptfont0=\tenrm
      \scriptscriptfont0=\sevenrm\rm}
\normalbaselines\rm}


\mathchardef\alpha="710B
\mathchardef\beta="710C
\mathchardef\gamma="710D
\mathchardef\delta="710E
\mathchardef\epsilon="710F
\mathchardef\zeta="7110
\mathchardef\eta="7111
\mathchardef\theta="7112
\mathchardef\kappa="7114
\mathchardef\lambda="7115
\mathchardef\mu="7116
\mathchardef\nu="7117
\mathchardef\xi="7118
\mathchardef\pi="7119
\mathchardef\rho="711A
\mathchardef\sigma="711B
\mathchardef\tau="711C
\mathchardef\phi="711E
\mathchardef\chi="711F
\mathchardef\psi="7120
\mathchardef\omega="7121
\mathchardef\varepsilon="7122
\mathchardef\vartheta="7123
\mathchardef\varrho="7125
\mathchardef\varphi="7127



\def\beginlinemode{\endmode
  \begingroup\parskip=0pt \obeylines\def\\{\par}\def\endmode{\par\endgroup}}
\def\beginparmode{\endmode
  \begingroup \def\endmode{\par\endgroup}}
\let\endmode=\par
{\obeylines\gdef\
{}}
\def\singlespace{\baselineskip=\normalbaselineskip}

\def\oneandahalfspace{\baselineskip=\normalbaselineskip
  \multiply\baselineskip by 3 \divide\baselineskip by 2}
\def\doublespace{\baselineskip=\normalbaselineskip \multiply\baselineskip by 2}

\nopagenumbers
\newcount\firstpageno
\firstpageno=2
\headline={\ifnum\pageno<\firstpageno{\hfil}\else{\hfil\elevenrm\folio\hfil}\fi}
\let\rawfootnote=\footnote		
\def\footnote#1#2{{\oneandahalfspace\parindent=0pt
\rawfootnote{#1}{#2}}}
\def\raggedcenter{\leftskip=4em plus 12em \rightskip=\leftskip
  \parindent=0pt \parfillskip=0pt \spaceskip=.3333em \xspaceskip=.5em
  \pretolerance=9999 \tolerance=9999
  \hyphenpenalty=9999 \exhyphenpenalty=9999 }
\def\dateline{\rightline{\ifcase\month\or
  January\or February\or March\or April\or May\or June\or
  July\or August\or September\or October\or November\or December\fi
  \space\number\year}}
\def\received{\vskip 3pt plus 0.2fill
 \centerline{\sl (Received\space\ifcase\month\or
  January\or February\or March\or April\or May\or June\or
  July\or August\or September\or October\or November\or December\fi
  \qquad, \number\year)}}


\hsize=6.5truein
\hoffset=0truein
\vsize=8.9truein
\voffset=0truein
\hfuzz=0.1pt
\vfuzz=0.1pt
\parskip=\medskipamount
\overfullrule=0pt	



\def\title			
  {\null\vskip 3pt plus 0.2fill
   \beginlinemode \doublespace \raggedcenter \bf}

\def\author			
  {\vskip 3pt plus 0.2fill \beginlinemode
   \singlespace \raggedcenter}

\def\affil			
  {\vskip 3pt plus 0.1fill \beginlinemode
   \oneandahalfspace \raggedcenter \sl}

\def\abstract			
  {\vskip 3pt plus 0.3fill \beginparmode
   \doublespace \narrower ABSTRACT: }

\def\summary			
  {\vskip 3pt plus 0.3fill \beginparmode
   \doublespace \narrower SUMMARY: }

\def\pacs#1
  {\vskip 3pt plus 0.2fill PACS: #1}

\def\endtitlepage		
  {\endpage			
   \body}

\def\body			
  {\beginparmode}		

\def\head#1{			
  \medskip\vskip 0.5truein	
  {\immediate\write16{#1}
   \raggedcenter \uppercase{#1}\par}
   \nobreak\vskip 0.25truein\nobreak}

\def\refto#1{$^{#1}$}		

\def\references			
  {\head{References}		
   \beginparmode
   \frenchspacing \parindent=0pt \leftskip=1truecm
   \parskip=8pt plus 3pt \everypar{\hangindent=\parindent}}

\gdef\refis#1{\indent\hbox to 0pt{\hss[#1]~}}	

\gdef\journal#1, #2, #3, 1#4#5#6{		
    {\sl #1~}{\bf #2}, #3 (1#4#5#6)}		

\def\refstylenp{		
  \gdef\refto##1{ [##1]}				
  \gdef\refis##1{\indent\hbox to 0pt{\hss##1)~}}	
  \gdef\journal##1, ##2, ##3, ##4 {			
     {\sl ##1~}{\bf ##2~}(##3) ##4 }}

\def\refstyleprnp{		
  \gdef\refto##1{ [##1]}				
  \gdef\refis##1{\indent\hbox to 0pt{\hss##1)~}}	
  \gdef\journal##1, ##2, ##3, 1##4##5##6{		
    {\sl ##1~}{\bf ##2~}(1##4##5##6) ##3}}

\def\endreferences{\body}

\def\figurecaptions		
  {\endpage
   \beginparmode
   \head{Figure Captions}
}

\def\endpage			
  {\vfill\eject}

\def\endpaper			
  {\endmode\vfill\supereject}


\def\ref#1{Ref.[#1]}			
\def\Ref#1{Ref.[#1]}			
\def\Refs#1{Refs.[#1]}			

\def\frac#1#2{{\textstyle{#1 \over #2}}}

\def\sla{\raise.15ex\hbox{$/$}\kern-.57em}
\def\leaderfill{\leaders\hbox to 1em{\hss.\hss}\hfill}
\def\twiddle{\lower.9ex\rlap{$\kern-.1em\scriptstyle\sim$}}
\def\bigtwiddle{\lower1.ex\rlap{$\sim$}}
\def\gtwid{\mathrel{\raise.3ex\hbox{$>$\kern-.75em\lower1ex\hbox{$\sim$}}}}
\def\ltwid{\mathrel{\raise.3ex\hbox{$<$\kern-.75em\lower1ex\hbox{$\sim$}}}}
\def\square{\kern1pt\vbox{\hrule height 1.2pt\hbox{\vrule width 1.2pt\hskip 3pt
   \vbox{\vskip 6pt}\hskip 3pt\vrule width 0.6pt}\hrule height 0.6pt}\kern1pt}

\newcount\eqnumber
\eqnumber=0
\def\Eqno#1{\global\advance\eqnumber by 1
    \expandafter\xdef\csname !#1\endcsname{\the\eqnumber}
    \eqno(\the\eqnumber)}
\def\Eqref#1{\csname !#1\endcsname}
\def\bR{I\kern-.4em R}
\def\bP{I\kern-.4em P}

\newcount\startpage

\catcode`@=11
\newcount\r@fcount \r@fcount=0
\newcount\r@fcurr
\immediate\newwrite\reffile
\newif\ifr@ffile\r@ffilefalse
\def\w@rnwrite#1{\ifr@ffile\immediate\write\reffile{#1}\fi\message{#1}}

\def\writer@f#1>>{}
\def\referencefile{
  \r@ffiletrue\immediate\openout\reffile=\jobname.ref%
  \def\writer@f##1>>{\ifr@ffile\immediate\write\reffile%
    {\noexpand\refis{##1} = \csname r@fnum##1\endcsname = %
     \expandafter\expandafter\expandafter\strip@t\expandafter%
     \meaning\csname r@ftext\csname r@fnum##1\endcsname\endcsname}\fi}%
  \def\strip@t##1>>{}}

\def\citeall#1{\xdef#1##1{#1{\noexpand\cite{##1}}}}
\def\cite#1{\each@rg\citer@nge{#1}}	

\def\each@rg#1#2{{\let\thecsname=#1\expandafter\first@rg#2,\end,}}
\def\first@rg#1,{\thecsname{#1}\apply@rg}	
\def\apply@rg#1,{\ifx\end#1\let\next=\relax
\else,\thecsname{#1}\let\next=\apply@rg\fi\next}

\def\citer@nge#1{\citedor@nge#1-\end-}	
\def\citer@ngeat#1\end-{#1}
\def\citedor@nge#1-#2-{\ifx\end#2\r@featspace#1 
  \else\citel@@p{#1}{#2}\citer@ngeat\fi}	
\def\citel@@p#1#2{\ifnum#1>#2{\errmessage{Reference range #1-#2\space is bad.}
    \errhelp{If you cite a series of references by the notation M-N, then M and
    N must be integers, and N must be greater than or equal to M.}}\else%
 {\count0=#1\count1=#2\advance\count1
by1\relax\expandafter\r@fcite\the\count0,%
  \loop\advance\count0 by1\relax
    \ifnum\count0<\count1,\expandafter\r@fcite\the\count0,%
  \repeat}\fi}

\def\r@featspace#1#2 {\r@fcite#1#2,}	
\def\r@fcite#1,{\ifuncit@d{#1}		
    \expandafter\gdef\csname r@ftext\number\r@fcount\endcsname%
    {\message{Reference #1 to be supplied.}\writer@f#1>>#1 to be supplied.\par
     }\fi%
  \csname r@fnum#1\endcsname}

\def\ifuncit@d#1{\expandafter\ifx\csname r@fnum#1\endcsname\relax%
\global\advance\r@fcount by1%
\expandafter\xdef\csname r@fnum#1\endcsname{\number\r@fcount}}

\let\r@fis=\refis			
\def\refis#1#2#3\par{\ifuncit@d{#1}
    \w@rnwrite{Reference #1=\number\r@fcount\space is not cited up to now.}\fi%
  \expandafter\gdef\csname r@ftext\csname r@fnum#1\endcsname\endcsname%
  {\writer@f#1>>#2#3\par}}

\def\r@ferr{\endreferences\errmessage{I was expecting to see
\noexpand\endreferences before now;  I have inserted it here.}}
\let\r@ferences=\references
\def\references{\r@ferences\def\endmode{\r@ferr\par\endgroup}}

\let\endr@ferences=\endreferences
\def\endreferences{\r@fcurr=0
  {\loop\ifnum\r@fcurr<\r@fcount
    \advance\r@fcurr by 1\relax\expandafter\r@fis\expandafter{\number\r@fcurr}%
    \csname r@ftext\number\r@fcurr\endcsname%
  \repeat}\gdef\r@ferr{}\endr@ferences}


\let\r@fend=\endpaper\gdef\endpaper{\ifr@ffile
\immediate\write16{Cross References written on []\jobname.REF.}\fi\r@fend}

\catcode`@=12

\citeall\refto		
\citeall\ref		%
\citeall\Ref		%
\citeall\Refs		%

\twelvepoint\doublespace

\title
Predictability of self-organizing systems
\author
S.L.~Pepke and J.M.~Carlson
\affil
Dept. of Physics,
University of California, Santa Barbara 93106
\abstract
We
study the predictability of large events in
self-organizing systems.
We focus  on a set of models which have  been studied
as  analogs of earthquake faults and fault systems,
and apply methods based on techniques which are of current interest in
seismology.
In all  cases we find   detectable correlations between
precursory smaller events and the large events  we aim to forecast.
We compare predictions based on  different patterns of precursory events
and find that  for all of the models
a new precursor based on the  spatial
distribution of  activity
outperforms more traditional measures
based on temporal variations in the   local   activity.

\pacs
91.30.Px, 05.45.+b, 02.50.+s, 05.20.-y

\endtitlepage

\body
\doublespace
\head{I. Introduction}
Self-organized criticality (SOC)  has received
considerable attention
over the past several years, as a possible means to explain
scaling behaviors observed in a broad class of nonequilibrium
systems including systems  in
geology, economics, and biology.\refto{BTW}
The theoretical prototype
is the sandpile model, in which sand is slowly
added to a pile and released in
instantaneous
avalanches of a wide range of sizes which  are triggered
when the height (or  slope or stress) locally  exceeds a
specified threshold.
Self-organized criticality refers to the particular case of
when  the system size
sets the cutoff for the largest events which are observed.
More generally,
self-organizing systems, whether critical or not, typically exhibit
scaling over some range of sizes and are thought to
evolve  so that
fluctuations in space and time are intrinsically coupled
by an underlying threshold dynamics.
There has recently been a considerable effort
to use self-organizing systems as simple   dynamical models of
seismic phenomena,\refto{BTW,OFC,CBO,CL} in part due to the
clear connection between earthquakes and    threshold dynamics.
Particular attention has been
paid to the
robust power-law scaling relation-- the Gutenberg--Richter law\refto{GR}--
relating the frequency of earthquakes to their  size.

An alternate direction of research
concerns the predictability of the
systems, which has important practical applications.
The basic approach in such an endeavor is to utilize
the available history of the system
to forecast future events.
Often one is most interested in predicting the largest events (e.g.~great
earthquakes),
and it is the largest events with which we will be concerned  here.
Of course, in a well defined deterministic system
precise knowledge about the present configuration of the system
will yield very good, if not perfect, prediction.
However, for many real systems  specific   equations
describing the detailed   evolutions of the systems are not known, and,
in addition,
the information on which the forecast must be based
is typically incomplete.
The basic prediction problem is, therefore, an inverse problem
in the sense that one wants to use  some information
such as  the time series of
events to infer something about the likely configuration
of the system which is, in most practical situations,
completely inaccessible to measurement.

For example,
earthquake catalogs  list the date, time, location,
and magnitude of detected events and thus  provide one possible source
from which one might hope to deduce information about
local stresses on a fault.
If correlations are detected they may
lead to measurable precursors
that are useful for forecasting.
While  certain  seismicity patterns have been recorded in catalogs prior
to some subset of the large earthquakes,
in most cases
the catalogs are too short
to  determine conclusively whether there
is a statistically significant correlation between
these patterns and large events.
Dynamical models of earthquake faults
can thus  be particularly useful in the context of the prediction
problem.
Study of models  allows
us to consider
catalogs of arbitrary
size, from which we can make
statistically significant statements about
both the intrinsic predictability of dynamical systems of this
type and the value of current prediction algorithms.

In this paper we address prediction issues in a variety
of self-organizing systems.  The algorithms that we
use are similar to, and clearly motivated by, the work
of Keilis-Borok, {\it et al.}\refto{M8} which we  describe briefly  below.
Our motivations for applying prediction algorithms to
a broad class of systems is to try to ascertain what classes of
precursory phenomena
are consistently  observed in all of the models.  The point here is
that no completely realistic dynamical model of  faults presently exists,
but  if the real system resembles in any substantive way the
threshold dynamics characteristic of the models, then precursors which
are observed in a broad class of  models may prove useful in real systems.
In the course of this endeavor we develop a new type of precursor
which is currently not in use in any form in the seismology community,
which performs particularly well in the models.
In addition,
an unanticipated result from our simulations is that
systems which have no apparent conservation law seem to
be more predictable than those with a conservation law.
Finally,  while our findings may have  important applications,
we would like to point out
the modest nature of our results.
The remaining open issues are extensive,
with the most striking and presumably difficult issue being
the development of an organized approach to deriving the
most efficient  precursors based on  limited spatio--temporal information
for a high dimensional dynamical
system.   In contrast,
here we address the issue of predictability in the context
of several fixed prescribed   precursors used in isolation.

\head{II. The models}

We  focus on a set of models which have been  suggested
as possible dynamical analogs of seismic phenomena which are
described briefly below.
We refer to the models as the
the Bak, Tang, and Wiesenfeld (BTW)  model,\refto{BTW}
the Olami,  Feder, and Christensen (OFC)  model,\refto{OFC}
the Chen, Bak, and Obukov  (CBO) model,\refto{CBO}
and the  Uniform Burridge and Knopoff (UBK) model.\refto{CL,BK}
More detailed descriptions of the individual models may be found
in the references cited above.

We begin with the UBK model
which satisfies a nonlinear wave equation
$${{\partial^2 U}\over{\partial t^2}}=
{{\partial^2 U}\over{\partial x^2}}-U -\phi(\dot U) +\nu t.\Eqno{eom}$$
Here $U(x,t)$ represents the relative displacement of  opposite sides
of a homogeneous fault as a function of position $x$ and time $t$.
The variable $\nu$ is the very slow shear rate driving the relative
motion of the plates,
and the key instability leading to chaotic behavior
is a velocity--weakening, stick--slip friction law $\phi(\dot U)$.
In the finite difference approximation, the model can be thought of as
a one-dimensional chain of blocks, which is pulled slowly
across a rough surface.

Unlike the UBK model all of the other
models ignore the details of inertial
dynamics and friction laws, and instead evolve according to
specified \lq\lq breaking rules," so that when the stress of a local
block exceeds a threshold it  relaxes according to some avalanche dynamics.
For each of these the system can be thought of as
a two dimensional\refto{2D}  square lattice of \lq\lq blocks" with open
boundary conditions.
In each case there is a
particular  rule  which specifies
the stress drop of the toppling site, the increases in stress of
other sites, and the net stress drop of  the system.

The BTW model is the original sandpile cellular automaton.
Of those we are considering,
it is the only model that is
driven stochastically: on each iteration of the automaton
the stress of a  randomly selected site  is increased by unity.
If that site is above a specified threshold stress it
initiates an avalanche
in which each toppling site loses four units of stress, giving
one unit to each neighbor (stress is dissipated  at the boundary),
and the cascade  continues until all sites are below threshold.

In   the OFC and CBO models
(as in the UBK model)
stress is increased uniformly across the
whole system.
The OFC model is  similar
to the BTW model in that equal stress is transferred to each neighbor
in each toppling,  however,
unlike the BTW model,
in the OFC model
the internal   dynamics
does not conserve stress.
Instead, in  each toppling
the stress of the toppling site is set to zero, and each neighbor receives
a fraction $\alpha<.25$ of the initial stress of the toppling site
(we will typically take $\alpha=.2$). Any remaining stress is
dissipated. This model has received considerable attention
as a possible non-conservative example of SOC, although
there has been some debate over
whether this model exhibits SOC in the thermodynamic limit.\refto{Grinstein}

Like the BTW model,
the CBO model does conserve stress away from the boundary. In
addition, while it is not driven stochastically, it is
does contain a stochastic element-- after each site topples,
its threshold stress is reset to a random value
chosen uniformly from $[0,1]$. In contrast, all of the other models
have a fixed uniform threshold.
Furthermore, among the models we are considering, the CBO model is
unique in that it
is the only one in which
the relaxation dynamics explicitly takes  place
using long range
elastic interactions, rather than by redistributing
stress only to neighboring sites.
Here the stress of the toppling site is set to zero and
the stress redistribution over the rest of the lattice is
viewed as that due to a dipole force at the toppling site
(hence falling off as $r^{-d}$).

While the BTW, OFC, and CBO models
clearly differ from one-another in certain important ways,
at least for the system
sizes considered here they  all
generate pure power law event size distributions:
$$P(s)={s^{-(b+1)}}\Eqno{GR}$$
(see Fig.~1),
analogous to the
Gutenberg--Richter law\refto{GR}
describing seismicity catalogs taken from
the entire Earth or
large regional  fault systems.
Thus we will refer to  these systems as
examples of SOC, with emphasis on \lq\lq criticality"
because the power law extends
from the smallest event size up to essentially the system size.

In contrast, the UBK model, while self-organizing,  is not critical.
The event size distribution consists of a power law describing
the small to moderate  events, and  excess  large
events, which cut off at some characteristic size, independent of the
system size\refto{CLST} for systems which are large enough (Fig.~1).
This  statistical distribution is analogous to what is
thought to apply to
individual faults, or narrow fault zones, where the largest events
appear to dominate the total slip,
occurring at a rate which exceeds the extrapolated rate of small to moderate
events.\refto{characteristic, Scholz}

\head{III. The prediction algorithm}

Our method of forecasting   resembles
the
algorithm M8 introduced
by
Keilis--Borok {\it et al.},\refto{M8}
which is currently being studied as a possible
means of using
worldwide seismic data sets to predict the largest earthquakes
in any given region.
The  M8 algorithm is based on the
hypothesis that  regional small scale seismicity may be used
to diagnose an upcoming large event.
The procedure is to first coarse grain the catalogs in space
and time, and then to measure certain precursors
in  these space--time windows.
Several   such
precursors are  used, and include   a variety of
measures based on the
activity $A$, which  within each space--time window is defined to be:
$A={\rm \# earthquakes}$  which are identified as main
shocks\refto{aftershocks}
 and
which are greater than or equal to some threshold size.
Note that $A$
is easily deduced from the  time series
of events in a region, and
large values of $A$ indicate
a regional temporal clustering of events.
More generally,
an effective precursor is one which will typically sustain
elevated (or depressed) values prior to a large event
relative to its average  value.

In
the Earth
no single measure  has  yet been identified which reliably
predicts all of the large events.
Instead the M8 algorithm  combines seven different
precursors  in
a  voting algorithm which is used to make predictions.
It is important to articulate
the  prediction goal defined by Keilis--Borok {\it et al.}
Instead of assigning some probability for an event
to occur at a specific place and time in the future,
the idea is simply to recognize certain seismicity patterns which
might indicate a time of increased probability, or \lq\lq TIP,"
for a large earthquake within a relatively large region in space
(the spatial windows are chosen to be an order of
magnitude larger than the event to be forecast).
In particular, if a fixed number of precursors exceed individual thresholds
in a region then the TIP is turned on.

The earliest applications of these
methods were based on existing data in real catalogs, and
it was found that in order to capture roughly 80\% of the large
events, approximately 20\% of the total space--time volume had to
sustain TIPs.
Efforts are currently underway
to evaluate the algorithms more thoroughly
by establishing systematic tests for forward prediction.\refto{Healy}
However, use of the algorithm  has  been    controversial
for a variety of reasons including
the intrinsic sensitivity  to  the inherent inaccuracies and
incompleteness of the  catalogs,\refto{Habermann}
and sensitivity of the algorithm
to features such as the initial placement of
test regions (the space--time windows), where  small adjustments
in the spatial positions of
the regions  and start dates of the catalogs
can easily cause the algorithm to miss some of the events.\refto{Minster}
Of course, given a perfectly accurate catalog of arbitrary length
(such as  can easily be generated for models),
the performance of these algorithms could easily
be assessed.
However, such catalogs are simply not  available, so that
the question of the  predictability of earthquakes,
as well as the development of  effective algorithms,
remain  open  and active areas of research.
Here we will assess
how well similar algorithms can be made to work
on a collection of dynamical  models.

We  consider a simplified version
of this prediction
algorithm, in which
precursors are considered individually.
We turn a TIP on when the precursor
exceeds a threshold,
and turn the TIP off when the precursor  falls below
the threshold.
By varying the threshold we  vary the total alarm time, and from this
we construct
a {\it success curve}\refto{Molchan} (Fig.~3), which plots the fraction of
events
predicted as a function of the fraction of the total space--time volume
occupied by
TIPs.

Comparison of the
success curves  allows us to compare the effectiveness of
different precursors as well as the predictability
(based on these precursors) of different models.
For a simple null test,
our results can also be compared
with the corresponding results for
purely random methods, in which TIPs are
issued completely arbitrarily.
In that case, events are predicted purely by chance, and the fraction which
are predicted successfully is simply given by the fraction of time the TIP
is on:
(\% predicted)=(\% alarm time), i.e.~the success curve is the diagonal  line.
In a purely random system no algorithm will perform better than this
method.
In our case,
this  gives us an operational definition of predictability  of models
using  a particular algorithm:
if a model
is predictable,  the success curve for some precursor
should  deviate  from this
line in a statistically significant manner.

It is not necessary for the success curve to  lie
above the diagonal line. In principal the curve could cross the line, or even
lie completely below it. Any deviation from the diagonal
is a sign that the precursor is  detecting
some correlation (or anticorrelation)  in the
system. However, in practice to obtain a success rate which is better,
rather than worse, than the corresponding results for random methods,
one must base  predictions on the complement of the original precursor
whenever the success curve lies below the diagonal line.\refto{combine}

Finally,
because of the limited amount of data which is available
in seismic catalogs
it is important to accompany any prediction
with an assessment of the associated confidence
level.
In contrast,
in our case we can generate catalogs that are arbitrarily long, and
thus obtain results to an arbitrarily high level of precision.
In particular, to verify that our success curves have converged to their
asymptotic limits in time, we generate independent curves for
a series of exponentially growing time intervals. This ability to check
for systematic convergence is especially useful
for models in which the predictability is marginal.

\head{IV. Results}

We begin with the UBK model.
A small segment
of the  catalog of events
is illustrated in Fig.~2.
For each event, a line is drawn through all the blocks which
slip.
While precursory small scale seismicity
is clearly correlated with  the epicenters of future
large events  (much more so for the model than for the earth)\refto{SCL}
it begins   on average after  half of the mean
recurrence interval between large events has elapsed,
so that from Fig.~2 alone it is not clear
how accurate predictions based    on this pattern  will be.

In \Ref{PCS} a detailed study of predictability of the UBK model
revealed that among a   set of precursors,
the two most effective
are activity $A$= \# earthquakes, and a new
measure   which is a better measure of  the development of
spatial correlations. This new measure, which we call
active zone size $AZS$,
is defined to be:
$AZS=$
\# blocks which have slipped (independent of the number of times)
in each space--time window.
With  $A$ we are
able to predict 90\% of the large events, with alarms which occupy 15\%
of the total space--time volume (Fig.~3), which corresponds to alarms
which occupy significantly   smaller time intervals than the average
duration of
small scale seismicity
prior to  large events\refto{PCS} in Fig.~2.
However,
the performance of  $AZS$ is even more impressive,
leading to successful predictions of 90\% of the large events
when alarms occupy only 8\% of the  space--time volume.\refto{counting}
In the UBK model the effectiveness of $AZS$ can be traced
to the fact that very little stress is relieved when a block slips
in a small event. Instead small events serve as markers that the region
is locally close to threshold.
While the two precursors are clearly not independent,
in contrast  to $A$, $AZS$ is a much
more direct measure of the size of the region that is near the
threshold for slipping, and thus ultimately
leads to the more direct assessment of the probability of a  large event.

The question remains: which precursors which worked well
for  the UBK model
are effective precursors for
other self-organizing  systems?
To address this, we next consider the SOC models.
Apart from some measurements of correlation functions between
events of similar size in the OFC
model\refto{OFC2} (which detected a tendency of large events to
cluster in time)
there has been little work to characterize the
SOC models in terms of the predictability.

As previously noted, the behavior of the UBK and SOC models
differ from one another substantially. In principle, for the SOC
models we could
coarse grain the catalogs of events in space and time
in a manner exactly
like that used for the UBK model.
However, because
in the SOC models the largest events span
essentially the entire system, and it is these largest events we wish to
forecast,
the most sensible choice is to define the spatial windows
to correspond to the entire system.
Furthermore, in these systems
the distinction between
small precursory events, and the large events which we attempt to
predict is no longer a sharp feature, as apparent in the statistical
distributions.
For that reason, we set a somewhat arbitrary lower cutoff for
events to predict
which  corresponds to the size where we estimate (by eye) that
finite size effects  first
become apparent (see Fig.~1).
Our preliminary estimates of
the predictability of
small and medium size events indicate
that in comparison the
largest events are
at least as predictable, and in most cases significantly more so than the
others.\refto{PC2}

The success curves for the SOC models, as well as our previous results
for the UBK model, are
illustrated in Fig.~3.
For all of the models
both  $A$ and $AZS$ yield success curves which deviate systematically from the
results obtained for random methods (the diagonal line),
and thus lead to some measurable predictability.
In each case
along essentially the entire success curve
$AZS$ gives the greater deviation, and hence
is more effective than $A$ as a precursor for a coming large event.
For the SOC models,
we find that  in most cases
these measures are in fact anticorrelated with  large events,
with success curves falling below the diagonal.
Whenever this is the case,
we plot   the complements of these measures,
i.e.~lack of activity
$\overline A$  (quiescence)
and lack of active zone size $\overline{AZS}$, because these are
the measures which would be used in practice to obtain a success rate
which is better than random methods.
When the complementary measure is used,
TIPs are turned on when our previous measures $A$ and $AZS$
take values below, rather than above, a specified threshold.
Of the SOC models, the  OFC   model is clearly most
predictable, generating a
success curve which is   comparable
to that of the UBK model.
In this case the most effective measure is
$\overline{AZS}$,
leading to  90\% events predicted with alarm times of
order 20\%.
Similarly, for the BTW and CBO models  $\overline{AZS}$
outperforms measures based solely on activity ($\overline A$  for the BTW
model,
and $A$ for the CBO model).  However,
compared to the UBK and  OFC models,   the gain
over purely random methods is significantly reduced.

In each case
there is  at least some   correlation between
small scale activity and coming large events,
suggesting (but by no means proving) that
self-organization may  have implications for the predictability
of real systems.\refto{Weissman}
The poor performance in the BTW and CBO models
indicates that the correlations need not be strong,
and may in real systems be sufficiently weak that external noise
or limited statistics
may mask their presence.
Recall that  both the BTW and CBO models contain  stochastic attributes,
which we expect play  important roles in limiting their predictability.
In contrast,
both the UBK  and  OFC models,  which exhibit the highest levels
of predictability,
are fully deterministic. However, the UBK and OFC models
have an additional common feature which differentiates them from the
BTW and CBO models.
Both the UBK and OFC models
do not satisfy a conservation law in the redistribution of total stress.
A more detailed study is necessary to fully
separate the relative roles of
deterministic dynamics and the lack of conservation
in determining the predictability of different systems,
even within our limited definition of predictability.
However, to address
this question
at least in part we have  considered
the predictability of the  OFC model
as the conservation parameter is varied.
In that case we  find that  the predictability
diminishes systematically as the level of conservation is increased,
although,
as illustrated in Fig.~4,
even when the OFC model is fully conservative its predictability
is clearly
greater than that illustrated in Fig.~3 for the BTW and CBO
models.\refto{OFC2,PC2}

In the SOC models the precursors based on  quiescence are typically
most effective
because, unlike  the UBK model, the stress
on a site is set to zero  each time a block slips, independent of the
event size.  Thus a lack of events
is more likely to signify that
the system is near  the slipping threshold.
Interestingly, in the Earth both increases and decreases
of seismicity have been observed prior to large events,\refto{Kanamori}
suggesting that perhaps  the UBK and SOC models may  all contain
some elements which are relevant to real faults.  Ultimately
it may be of interest to incorporate the more  complete
dynamical treatment of individual faults present in the UBK model
into an SOC--type model which describes fault zones.

In both the UBK and SOC models,
large events  involve large spatial regions, so that in order
for a large event to occur, the system must be
near threshold across a relatively large region in space.
In all of the  models considered here,
methods based on $AZS$
are more effective than those based on $A$,
since $AZS$ provides  the more direct measure of the development of
such  regions.
The strong performance of this new precursor
is particularly noteworthy  because
such measures are not currently being used  quantitatively
in earthquake prediction algorithms such as M8.
Certain tendencies towards clustering in space and time
have been noted,\refto{Kagan}
and are the phenomenological basis of these algorithms.
However, the precursors  which are  used
are based on measures  such as $A$
which within a given region
track the development of temporal correlations.
In such measures, spatial correlations are only accounted for in the
most primitive way-- i.e.~in the initial definition of the spatial
window.
In the Earth  including  precursors which
measure  the development of geometric spatial correlations
is complicated by the
inhomogeneity of fault networks and the difficulties associated
with accurately locating slip. Nonetheless,  a
box counting algorithm, in which the current spatial windows are
coarse grained, and a count is made
of regions exhibiting seismicity above or below some background level,
may be adequate  to  measure the analog of
$AZS$ or  $\overline{AZS}$.
It would be of significant interest
to assess the  performance  of such a  measure
in comparison to activity based
precursors in the Earth.

\noindent
$Acknowledgements:$
We thank  Volodya Kossobokov, James Langer,
and Glen Swindle for useful comments on the
manuscript.
This work  was supported by   the Alfred P.
Sloan and David and Lucile Packard
Foundations, NSF grant DMR-9212396, and
an INCOR grant from  LANL.


\references
\singlespace

\refis{counting} This method of obtaining the success curve
for the UBK model differs from that used in \Ref{PCS}
where success curves were obtained individually for each of the
spatial regions, and then averaged to obtain the cumulative results.
Here we consider alarms as they apply to the total space--time
volume to obtain a somewhat more direct comparison to the
SOC models.

\refis{combine} A combination of the original measure and the
complement measure will be most effective near places where the
success curve crosses the diagonal line.

\refis{Habermann} R.E.~Habermann, {\it Tectonophysics} {\bf 193},
277 (1991).

\refis{Molchan}  G.M.~Molchan,
{\it Tectonophysics} {\bf 193}, 267 (1991).

\refis{Kagan}
Y.Y.~Kagan,
{\it Geophys. J. R. Astr. Soc.} {\bf 71}, 659
(1982).

\refis{aftershocks}  A main shock is defined to be the largest event
in a sequence of earthquakes, consisting of a main shock followed
by aftershocks, and possibly preceded by foreshocks.
Various different  algorithms exist for
associating aftershocks with a particular main shock
in terms of their relative separation
in space and time from the main shock.
While,
one of the most prominent
phenomena associated with  real earthquakes
are their tendency to generate  long sequences of aftershocks,
unfortunately we cannot consider
aftershocks in this study because none of the models contains the time
scales associated with aftershock sequences.



\refis{Scholz}
C. Scholz,  {\it The mechanics of earthquakes and faulting},
Cambridge University Press, New York (1990).


\refis{Weissman} K. O'Brian,  M. Weissman,
{\it Phys. Rev. A} {\bf 46}, R4475 (1992).





\refis{GR}
B. Gutenberg and C. Richter, {\it Seismicity of the Earth
and Related Phenomena}, Princeton University Press, Princeton, NJ  (1954).


\refis{OFC} Z. Olami, H. Feder,  K. Christensen,
{\it Phys. Rev. Lett.} {\bf 68}, 1244 (1992).

\refis{OFC2} K. Christensen,  Z. Olami,
{\it J. Geophys. Res.} {\bf 97}, 8729 (1992).



\refis{PC2} Issues associated with optimization will be
discussed more fully in S.~Pepke and J.~Carlson, (in preparation).

\refis{2D}
The 2d UBK model differs little from the 1d case.
J. Carlson,
{\it Phys. Rev. A} {\bf 44},
6226 (1991).

\refis{BTW}
P. Bak, C. Tang, K. Wiesenfeld,
{\it Phys. Rev. Lett.} {\bf 59}, 381 (1987).

\refis{CLST}
J. Carlson,  J. Langer, B. Shaw, C. Tang, {\it Phys. Rev. A} {\bf 44},
884 (1991).

\refis{BK}
R. Burridge,  L. Knopoff,  {\it Bull.
Seismol. Soc. Am.} {\bf 57}, 3411 (1967).

\refis{CL}
J. Carlson, J. Langer,
{\it Phys. Rev. Lett.} {\bf 62}, 2632 (1989).



\refis{PCS}
S. Pepke, J. Carlson, B. Shaw,
\lq\lq Prediction of Large Events
on a Dynamical Model of a Fault,"
to appear in {\it J. Geophys. Res}).

\refis{CBO}
K. Chen,  P. Bak,  S. Obukov,
{\it Phys. Rev. A} {\bf 43}, 625 (1991).

\refis{characteristic}
See, e.g.,
D.P. Schwartz,  and K.J. Coppersmith,
{\it J.
Geophys. Res., 89}, 5681 (1984).

\refis{Healy}
J. Healy, V. Kossobokov,  J. Dewey,
{\it U. S. Geo. Sur. Open File Rep.  92-401} (1992).

\refis{Kanamori}
H. Kanamori,
{\it Earthquake Prediction, an International Review},
Maurice Ewing Series, Vol. 4,
Amer. Geo. Un., Washington, D.C., 1-19 (1981).

\refis{M8}
V. Keilis-Borok,  V. Kossobokov,
{\it Phys. Earth Planet. Inter.} {\bf 61},
73 (1990).

\refis{Grinstein}
J. Socolar,  G. Grinstein,  C.~Jayaprakash, {\it Phys. Rev. E}
{\bf 47}, 2366 (1993).

\refis{Minster}
J.-B. Minster,  N.P. Williams,  {\it EOS Transactions, AGU 1992 Fall
Meeting} {\bf 73}, 366 (1992).



\refis{SCL}
B. Shaw, J. Carlson, and J. Langer,
{\it J. Geophys. Res., 97}, 479 (1992).

\endreferences

\vfill\eject

\centerline{FIGURES}
\singlespace

\noindent
(1). Event size distributions $P(s)$ vs.~$s$ for the (a) UBK,
(b) OFC, (c) BTW, and (d) CBO models.
In each case $s$ is a measure of the size of the event--
the integrated slip (seismic moment) for the UBK model, and the number of
sites which topple  for the others. In each
case we attempt to predict events with
$s\ge\tilde s$, where $\tilde s$ is some characteristic size
(in each case $\tilde s$ is marked with an arrow).
The UBK model exhibits a sharp distinction between small
($s< \tilde s$) and large ($s\ge\tilde s$) events,
while the others exhibit
power laws cut off at  $s\approx \tilde s$
(determined by  finite size effects).
In these cases,
to exactly determine this crossover length a careful study of finite size
effects must be performed.  For the purposes of qualitative
comparisons, it is sufficient to roughly estimate $\tilde s$
as we do here.
We take:
$N=8192$, $\sigma=.01$, $\alpha=3$, and $\xi/a=10$
for the UBK
model,\refto{PCS}
system sizes $32\times 32$
for the other models, and
$\alpha=.2$ for the OFC model.\refto{OFC}

\noindent
(2). A small sample catalog  as a
function of space $x$  (block number)
and time $t$
in the UBK model.
A line segment marks
blocks which slip in each event,
and a cross  marks  the epicenter of each large event, which is clearly
correlated with the small scale seismicity.
The box corresponds to a space--time window within which
$A$ and $AZS$
are evaluated.

\noindent
(3). Success curves
for the (a)  UBK, (b) OFC, (c) BTW, and
(d) CBO models.
For  each model, the  best
spatial measure
($AZS$ or $\overline{AZS}$)
leads to more precise predictions
than the best temporal measure
($A$ or $\overline A$).
In each case, results for the complement measure (e.g.~$A$ vs.~$\overline A$)
are obtained by  a reflection of the curve across the diagonal.
In
the UBK model,
we predict  epicenters of large events,
and  optimal spatial windows are  less than the size of the large event
as shown in Fig.~2.
We coarse grain the system
into many  overlapping spatial regions,
and then obtain the success curve by calculating the fraction
of events successfully predicted by some   window containing the event
vs.~the total fraction of the
space--time volume which is occupied bt TIPs.\refto{counting}
In the other models, the spatial windows are taken to be the entire system,
and the goal is to  predict the   large event.
We have performed a crude optimization to select the time windows
within which the precursors are evaluated. The values
correspond to
$\Delta t=.1$ for
the UBK model (where the spatial windows were taken to be 213 blocks),
33 net grains added during the time window for the BTW model,
.15 net stress added per site for the OFC model,
and .007  net stress added per site in the measurement
of $\overline{AZS}$ and .015
net stress added per site  in the measurement
of $A$ for the CBO model.

\noindent
(4). The deterministic and fully conservative OFC model.
Here
$\alpha=.25$
whereas the results presented in Figs.~1 and 3 correspond to the dissipative
case $\alpha=.2$. Figure (a) illustrates
the event size distribution, with $\tilde s$ marked as in Fig.~1,
and (b) illustrates the success curve, analogous to Fig.~3.
Compared to the nonconservative case (Fig.~1b), here the maximum event size
is significantly increased, and  compared to Fig.~3b
the predictability is suppressed. As in Fig.~3, here we have crudely optimized
over time windows to select a window which corresponds to
.05 net stress added per site.

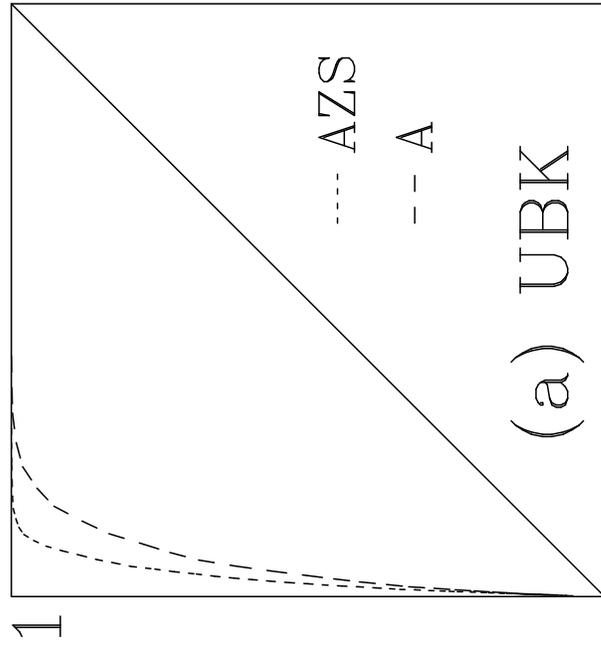
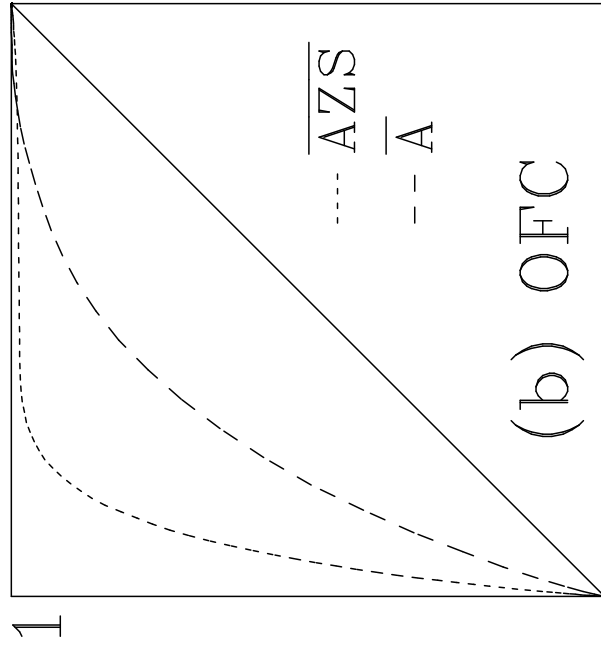
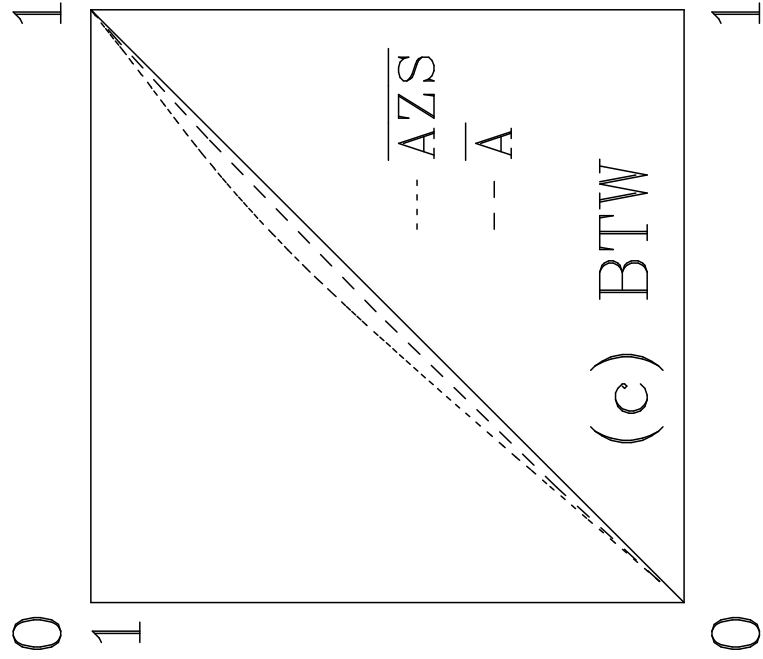
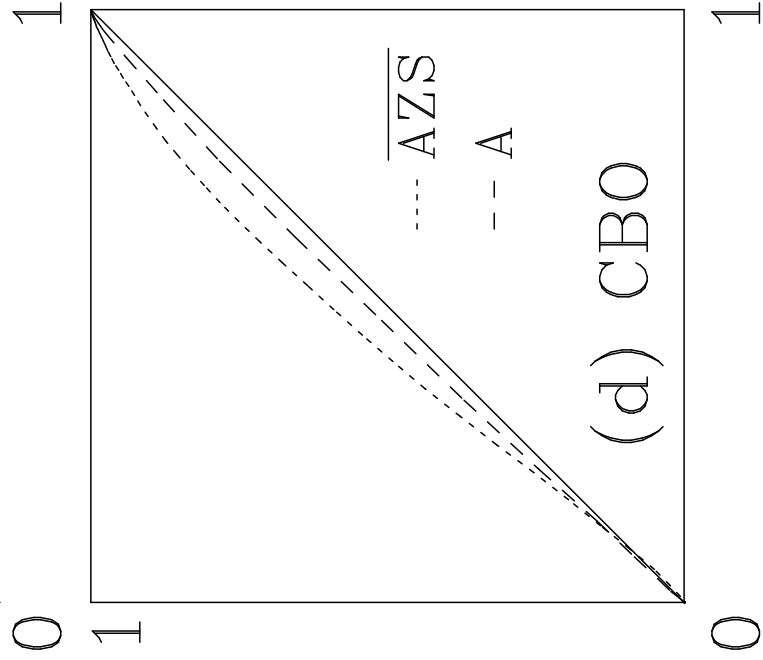
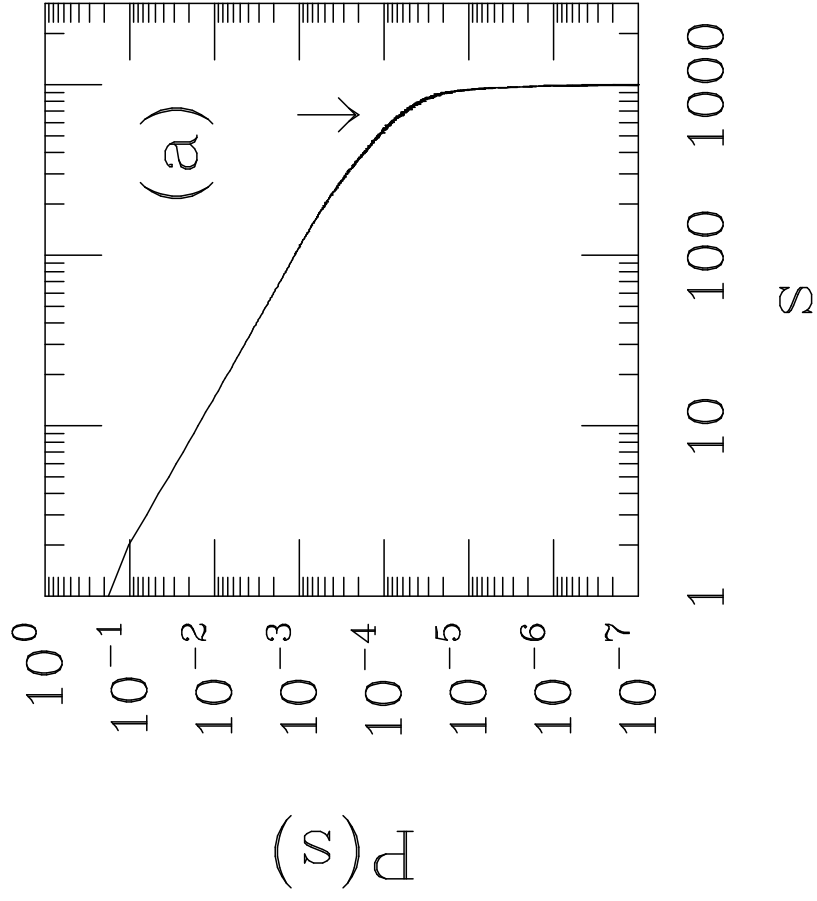
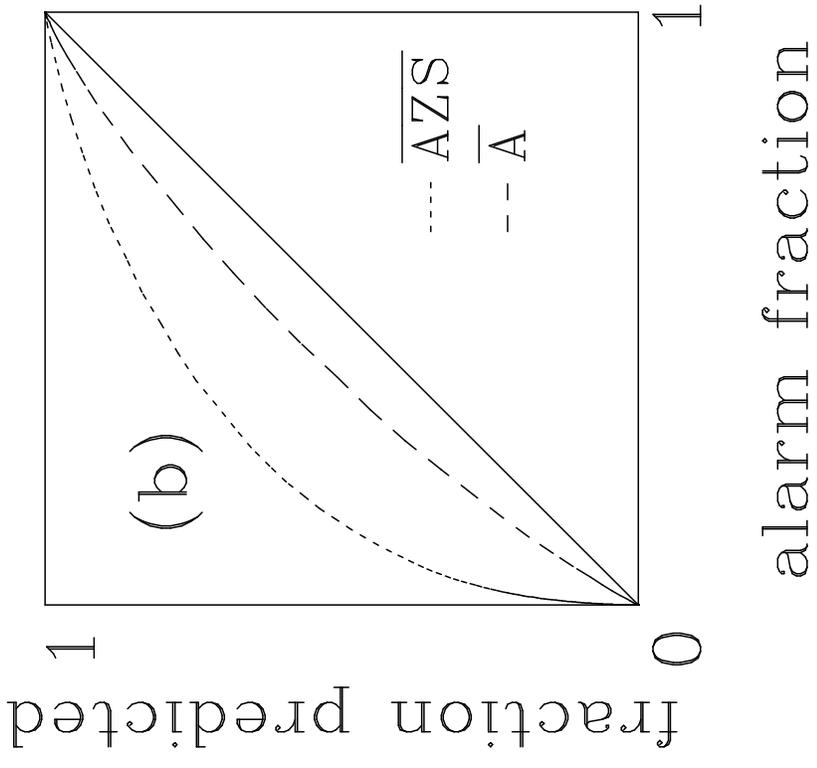
\end